\documentstyle[twocolumn,psfig,prl,aps]{revtex}
\makeatletter
\def\eqalign#1{\null\vcenter{\def\\{\cr}\openup\jot\m@th
  \ialign{\strut$\displaystyle{##}$\hfil&$\displaystyle{{}##}$\hfil
      \crcr#1\crcr}}\,}
\makeatother
\begin{document}

\twocolumn[\hsize\textwidth\columnwidth\hsize
\csname@twocolumnfalse\endcsname

\title{Bloch electron in a magnetic field and the Ising model}

\author{I. V. Krasovsky\cite{addr}}
\address{
Max-Planck-Institut f\"ur Mathematik in den Naturwissenschaften\\
Inselstr. 22--26, D-04103, Leipzig, Germany\\
and\\
B.I.Verkin Institute for Low Temperature Physics and Engineering\\
47 Lenina Ave., Kharkov 310164, Ukraine}

\maketitle

\begin{abstract}
The spectral determinant $\det(H-\varepsilon I)$ of the Azbel-Hofstadter 
Hamiltonian $H$ is related to Onsager's partition function of the 
2D Ising model for any value of 
magnetic flux $\Phi=2\pi P/Q$ through an elementary cell,
where $P$ and $Q$ are coprime integers. The band 
edges of $H$ correspond to the critical temperature of the Ising model;
the spectral determinant at these (and other points
defined in a certain similar way) is independent of $P$. 
A connection of the mean of Lyapunov exponents 
to the asymptotic (large $Q$) bandwidth is indicated.
\end{abstract}

 ] 

Although the problems of an electron in a constant magnetic field and 
that of an electron on a periodic lattice were solved in the early days 
of quantum mechanics, the case where a magnetic field and a lattice are
present simultaneously still defies adequate understanding. The simplest 
model for this case is commonly referred to as the 
Hofstadter or Azbel-Hofstadter 
model \cite{A,Hof}. The corresponding Hamiltonian describes an electron 
on a two-dimensional (2D) $N\times N$ square lattice with nearest-neighbour 
hopping subject to a perpendicular uniform magnetic field:
\begin{equation}
\eqalign{(H\psi)_{n_x,n_y}=
\psi_{n_x-1, n_y}+\psi_{n_x+1, n_y}+\\
\lambda e^{-in_x\Phi}\psi_{n_x, n_y-1}+
\lambda e^{in_x\Phi}\psi_{n_x, n_y+1},\\
n_x,n_y=0,1,\dots,N-1,}\label{1}
\end{equation}
where $\Phi=2\pi P/Q$ is the flux through an 
elementary cell (measured in units of the elementary flux),
$P$ and $Q$ are coprime (that is they do not have a common divisor 
other than 1) positive integers, 
$\lambda=t_1/t_2\ge 0$ is a ratio of hopping amplitudes in the x and y
directions.
We impose the periodic boundary conditions 
($\psi_{N, n_y}=\psi_{0, n_y}$, 
$\psi_{-1, n_y}=\psi_{N-1, n_y}$, similarly for $n_y$ ).
Usually, the Hamiltonian (\ref{1}) is defined
on an infinite lattice at the outset; in the present paper, however, we shall
deal first with a finite $N$ and take the limit $N\to\infty$ later on. 

This Hamiltonian and the related one-dimensional operator were studied
in many works (see  \cite{W,LJ,KS} for reviews).
Nevertheless, the quantitative description of the spectrum remains to
a large extent an open problem. 
Hamiltonian (1) is related to that of a quantum particle on a line in
the presence of two periodic potentials. 
It is the effect of incommensurability of the periods (when
$\Phi/2\pi$ is irrational or, in other words, $P,Q\to\infty$) leading to the
fractal structure of the spectrum that makes the problem both
difficult and interesting. Its solution would be an essential
contribution to the theory of fractals, would be important for studies 
of localization-delocalization phenomena, the quantum Hall effect, as well 
as purely for number theory and functional analysis.

It was noticed by Wiegmann and Zabrodin \cite{WZ} that the model is 
related to integrable systems and the algebra $U_q(sl_2)$. This was
generalized in \cite{FK}.
In the present paper, we find a different type of relation to integrable 
systems. We shall see that the spectral determinant 
$\det(H-\varepsilon I)$  for any $\Phi=2\pi P/Q$ is simply mapped onto 
Onsager's partition function of the 2D Ising model 
(formulas (\ref{pf2})--(\ref{eqv})).
This relation is, in a sense, complementary to the one established in
\cite{WZ,FK}: we shall explain this later in the text.

In the present paper, we remove
part of the mystery about the Thouless conjecture on the total bandwidth.
It is well known that the spectrum of $H$ for an infinite lattice (on $l_2(Z)$)
consists of $Q$ intervals (bands). When $Q\to\infty$, the total
width $W$ of the bands for $\lambda\ne 1$ is known \cite{AMS} to approach
$4|1-\lambda|$ exponentially fast in $Q$; for $\lambda=1$,
it is of order $1/Q$ \cite{L}. Thouless formulated a 
conjecture \cite{T83,Tcmp} that for $\lambda=1$ the total width
$W\sim 32G/\pi Q$ as $Q\to\infty$, where 
$G=1-1/3^2+1/5^2-1/7^2\dots$ is Catalan's
constant. (Notation $A\sim B$ means, henceforth, that $A/B$ tends to 1 
in the limit.) This result was derived only for $P=1$, $P=2$
so far \cite{Tcmp,T2,Watson,HK} but is supposed to hold in the 
general case (even when $P/Q$ tends to {\it any} nonzero limit) for which it is
supported by extensive numerical data; thus, the bandwidth is supposed to 
have an interesting universality property with respect to $\Phi$.
We shall generalize this conjecture and recast it 
into a simpler type of statement (expression (\ref{wg})) by relating 
the quantity
$\gamma(\varepsilon)=\lim_{N\to\infty}(1/ N^2)\ln|\det(H-\varepsilon I)|$ 
(which is interpreted as a mean of Lyapunov exponents)
to the bandwidth. In particular, for any band edge $\varepsilon_{e_i}$, 
{\it any} coprime $P$, $Q$, and $\lambda=1$, we get an exact result:
$\gamma(\varepsilon_{e_i})=4G/\pi Q$. This differs from the supposed
asymptotic formula for the total bandwidth only by the factor~8.

First, let us use the translational symmetry of $H$ in the x and y directions.
Substitute
$\psi_{n_x, n_y}=e^{ik_yn_y}\mu_{n_x}$, $k_y=2\pi k/N$, $k=0,1,\dots,N-1$
into (1) to get $H=\oplus_{k=0}^{N-1}H_k$, the eigenvalue equation for $H_k$
being
\begin{equation}
\eqalign{
(H_k\mu)_n=\\
\mu_{n-1}+\mu_{n+1}+2\lambda\cos\left(\frac{2\pi P}{Q}n+
\frac{2\pi k}{N}\right)\mu_n=
\varepsilon\mu_n,\\
n=n_x=0,1,\dots,N-1.}\label{hk}
\end{equation}
When $n$ is allowed to range from $-\infty$ to $\infty$, the corresponding
$H_k$ on $l_2(Z)$ is called the almost Mathieu operator.

Let us assume that $N$ is divisible by $Q$ and substitute 
$\mu_{j+Ql}=e^{i\omega l}\xi_j$, $j=0,1,\dots, Q-1$,
$\omega=\frac{2\pi m}{N/Q}$, $m=0,1,\dots, N/Q-1$ into (\ref{hk}).
We get $H_k=\oplus_{m=0}^{N/Q-1}H_{km}$,
\begin{equation}
\eqalign{
(H_{km}\xi)_j=\xi_{j-1}+\xi_{j+1}+2\lambda\cos\left(\frac{2\pi P}{Q}j+
\frac{2\pi k}{N}\right)\xi_j\\
j=0,1,\dots,Q-1;\qquad \xi_Q=e^{i\omega}\xi_0,
\qquad \xi_{-1}=e^{-i\omega}\xi_{Q-1}.}
\label{hkm}
\end{equation}

\begin{figure}
\centerline{\psfig{file=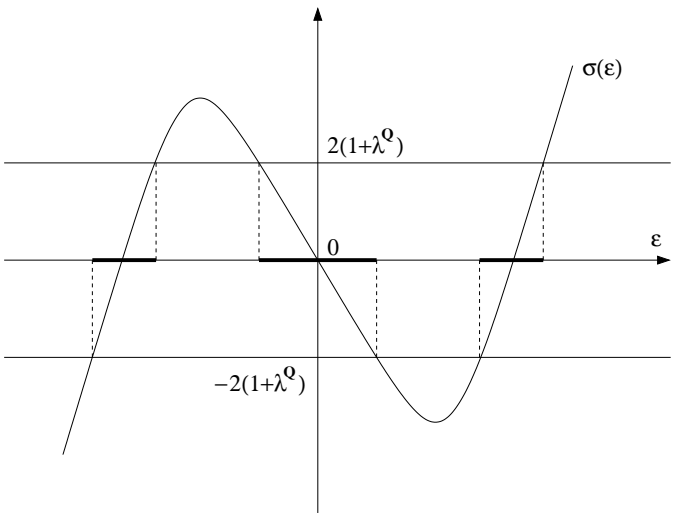,width=3.5in,angle=-90}}
\vspace{0.1cm}
\caption{
The polynomial $\sigma(\varepsilon)$ is sketched for $Q=3$.
The bands of the spectrum are shown by thick lines.}
\label{fig1}
\end{figure}

The Chambers formula \cite{Chambers} gives the dependence of the spectral 
determinant (characteristic polynomial) of $H_{km}$ on $k$ and $m$, namely:
\begin{equation}
\eqalign{
\det(H_{km}-\varepsilon I_{Q\times Q})=\\
(-1)^Q\left(\sigma(\varepsilon)-2\lambda^Q
\cos\frac{2\pi k}{N/Q}-2\cos\frac{2\pi m}{N/Q}\right),}\label{ch}
\end{equation}
where $\sigma(\varepsilon)$ is a polynomial of degree $Q$ which depends on 
$\lambda$ and $\Phi$, but not on $k$ and $m$. It is easy to verify (\ref{ch})
by considering the matrix given by (\ref{hkm}) and the one obtained after
the substitution $\xi_j=\sum_{l=0}^{Q-1}\exp 2\pi i(jlP/Q+jm/N+lk/N)
\xi^{'}_l$. 
It is the expression (\ref{ch}) which implies 
that the spectrum of $H$ in the limit $N\to\infty$ consists exactly of $Q$ 
bands: the image of the interval 
$[-2(1+\lambda^Q),2(1+\lambda^Q)]$ under the inverse of the transform 
$\sigma=\sigma(\varepsilon)$ (see Fig. \ref{fig1}). 

Thus,
\begin{equation}
\eqalign{
\det(H-\varepsilon I)=\prod_{k=0}^{N-1}\prod_{m=0}^{N/Q-1}
\det(H_{km}-\varepsilon I_{Q\times Q})=\\
(-1)^{N^2}\prod_{k,m=0}^{N/Q-1}\left(\sigma(\varepsilon)-2\lambda^Q
\cos\frac{2\pi k}{N/Q}-2\cos\frac{2\pi m}{N/Q}\right)^Q.}\label{det}
\end{equation}

In the limit of infinite lattice ($N\to\infty$), let us replace $2\pi k/N$ in 
(\ref{hk}) by a continuous parameter $\theta$, denote $H_k$ by $H_\theta$,
and consider the mean $\gamma(\varepsilon)$ of Lyapunov exponents over all 
real $\theta$. By virtue of the 3-term recursion (\ref{hk}), any $\mu_n$ is
obtained from initial conditions $\mu_0$, $\mu_1$.  
The Lyapunov exponent $\gamma_\theta(\varepsilon)$
corresponding to $H_\theta$ describes the exponential rate of growth (or decay)
of a solution to the equation $(H_\theta\mu)_n=\varepsilon\mu_n$, 
$n=0,1,\dots$. More precisely, 
\begin{equation}
\gamma_\theta(\varepsilon)=\lim_{n\to\infty} n^{-1}
\max_{\mu_0,\mu_1;\; \mu_0^2+\mu_1^2=1}
\ln(\mu_n^2+\mu_{n+1}^2)^{1/2}.
\end{equation}
Note that the difference between $H_\theta$ defined in the space of
square-summable sequences with indices $n=0,1,\dots$ and the almost Mathieu 
operator corresponding to $n=\dots,-1,0,1,\dots$ is that the spectrum of the 
latter is doubly degenerate. However, the normalized to unity density
of states  $\rho_\theta(x)$ is the same in both cases.
According to the Thouless formula \cite{le}, 
$\gamma_{\theta}(\varepsilon)=\int\ln|\varepsilon-x|\rho_\theta(x)dx$.
Using this formula and (\ref{det}), we get for the mean of 
$\gamma_\theta(\varepsilon)$ over $\theta$
(to ensure that the determinant is nonzero
we assume that $\varepsilon$ has an imaginary part; after taking 
the limit $N\to\infty$ we can let $\Im\varepsilon\to 0$):
\begin{equation}
\eqalign{
\gamma(\varepsilon)=\lim_{N\to\infty}{1\over N^2}\ln|\det(H-\varepsilon I)|=\\
{1\over \pi^2 Q}\int_0^\pi\int_0^\pi
\ln|\sigma(\varepsilon)-2\lambda^Q\cos x-2\cos y|dxdy.}\label{pf1}
\end{equation}

On the other hand, Onsager's partition function $Z$ for the 2D 
Ising model on a square $N\times N$ lattice satisfies \cite{Ons}:
\begin{equation}
\eqalign{
\lim_{N\to\infty}{1\over N^2}\ln Z-{1\over2}\ln(2\sinh a')=\\
{1\over 2\pi^2}\int_0^\pi\int_0^\pi\ln\left|2\frac{\cosh a \cosh a'}{\sinh a'}-\right.\\
\left. 2\frac{\sinh a}{\sinh a'}\cos x-2\cos y\right|dxdy.}\label{pf2}
\end{equation}

Here $a=2J/T$, $a'=2J'/T$, where $T$ is the temperature; $J$, $J'$,
interaction constants in $x$ and $y$ direction.
We now set $\sinh a/\sinh a'=\lambda^Q$ (hence 
$a'={\rm arcsinh}\,(\lambda^{-Q}\sinh a)$);
\begin{equation}
\sigma(\varepsilon)=2\frac{\cosh a \cosh a'}{\sinh a'}=
2\lambda^Q\coth a\sqrt{1+\lambda^{-2Q}\sinh^2 a};\label{sigT}
\end{equation}
to obtain the equivalence (as $N\to\infty$) for any $\varepsilon$
and coprime $P$ and $Q$:
\begin{equation}
|\det t_2(H-\varepsilon I)|\sim Z^{2/Q},\label{eqv}
\end{equation}
where $t_2^Q=2\sinh a'$.

A simple analysis of (\ref{sigT}) shows that for any real $T$,
$|\sigma(\varepsilon)|\ge 2(1+\lambda^Q)$. The minimum of 
$|\sigma(\varepsilon)|$
as a function of $T$,  $|\sigma(\varepsilon)|=2(1+\lambda^Q)$,
corresponds to the critical temperature $T_c$ of the Ising model
(at $T_c$ the argument of the
logarithm in (\ref{pf1}) and (\ref{pf2}) vanishes at an integration limit).
On the other hand, the energies $\varepsilon_{e_i}$ at which
$|\sigma(\varepsilon_{e_i})|=2(1+\lambda^Q)$ are the band edges of 
$\lim_{N\to\infty} H$.

It is well known \cite{dimer}
that $Z$ at $T_c$ equals the square of the partition 
function $Z_D$
of dimers on the lattice if $r_1=\sqrt{2\sinh a}$ and $r_2=\sqrt{2\sinh a'}$
are interpreted as activities of dimers in $x$ and $y$ directions.
Let us identify $r_1^2=t_1^Q$ and $r_2^2=t_2^Q$ (recall that $t_1$ and 
$t_2$ are hopping amplitudes in $x$ and $y$ directions, and 
$\lambda=t_1/t_2$).
One result of Lieb and Loss \cite{LL} says that 
$\det\widetilde{H}\sim Z^2_D$ as $N\to\infty$ for the flux $\Phi=\pi$,
that is for $P=1$, $Q=2$. Here $\widetilde{H}=t_2 H$ --- a more symmetric 
form of the Azbel-Hofstadter Hamiltonian.
Since it is known that $\varepsilon=0$ is a band edge for $Q$ even, we have 
from (\ref{eqv}) $\det\widetilde{H}\sim Z^{4/Q}_D$,
which is a generalization of the mentioned result to any 
coprime $P$ odd, $Q$ even. 

In the works \cite{WZ,FK} the Bethe ansatz equations are
formulated for the zeros of the characteristic polynomial of $H_{km}$, 
in other words, for the roots of the equation
$\sigma(\varepsilon)={\mathrm const}$. Thus \cite{WZ,FK} are to 
do with the ``internal'' structure of $\sigma(\varepsilon)$, whereas
in our formalism $\sigma(\varepsilon)$ enters as an ``indivisible''
object and our results come from what appears (at least at first
sight) ``external'' to $\sigma(\varepsilon)$ structure of the problem:
the translational symmetry and the Chambers formula. 

Henceforth, we only consider the limit of infinite $N$. 

Let us simplify expression (\ref{pf1}) (note an important fact that 
$\gamma$ depends on $\varepsilon$ and $P$ only through 
$\sigma(\varepsilon)$ !).
Because of the obvious equality $\gamma(\sigma,\lambda)=
\ln\lambda+\gamma(\sigma/\lambda^Q,1/\lambda)$,
it is sufficient to consider only the case $0\le\lambda\le 1$.
In this case, taking the integral over $y$ in (\ref{pf1}) we get:
\begin{equation}
\eqalign{
\gamma(\sigma)=\\
\cases{
\frac{1}{\pi Q}\int_0^\pi{\mathrm arccosh}\,(|\sigma|/2+\lambda^Q\cos x)dx,\cr
\qquad |\sigma|\ge 2(1+\lambda^Q);\cr
\frac{1}{\pi Q}\int_0^{\arccos\{(2-|\sigma|)/2\lambda^Q\}}
{\mathrm arccosh}\,(|\sigma|/2+\lambda^Q\cos x)dx,\cr
\qquad 2(1+\lambda^Q) >|\sigma|> 2(1-\lambda^Q);\cr
0, \qquad |\sigma|\le 2(1-\lambda^Q).}}\label{Le}
\end{equation}
The fact that $\gamma(\varepsilon)$ is zero on the image of the interval
$\sigma\in [-2(1-\lambda^Q),2(1-\lambda^Q)]$ is in agreement with the general
argument: this image is an intersection of the intervals of the 
spectra of the operators $H_\theta$. The generalized eigenfunction of any
$H_\theta$ is just a Bloch wave on a periodic lattice with $Q$ atoms in 
an elementary cell. Therefore, the  
Lyapunov exponents $\gamma_\theta(\varepsilon)$ are zero on this set. 
By definition, $\gamma(\varepsilon)$ is a mean of these exponents. 

Let us remark that rather than considering $\det(H-\varepsilon I)$,
it is possible to derive equations (\ref{Le}) making use of some 
known properties (see, e.g., \cite{N}) of the monodromy operator.
 
Now we can note that in the limit $Q\to\infty$, $\sigma$ fixed, 
$\gamma(\sigma)$ tends to the following Lyapunov exponent
{\it on the spectrum} of $H$:
$\gamma(\varepsilon)\to 0$ if $\lambda\le 1$,
$\gamma(\varepsilon)\to\ln\lambda$ if $\lambda > 1$,
in accordance with a statement of Aubry and Andr{\'e} \cite{AA}.
(Naturally, we obtain the same asymptotic result for any individual
$\gamma_\theta(\varepsilon)$.)

For $T_c$ ($|\sigma|=2(1+\lambda^Q)$), let us represent (\ref{Le}) 
in another form by reducing the integral to $\int_0^{\pi/2}
{\mathrm arcsinh}\,(\lambda^{Q/2}\cos x) dx$ and then using
differentiation w.r.t. the parameter. We get for any $\lambda>0$:
\begin{equation}
\eqalign{
\gamma(\varepsilon_{e_i})=
\frac{4}{\pi Q}\int_0^{\lambda^{Q/2}}\frac{\arctan x}{x}dx=\\
\ln\lambda+\frac{4}{\pi Q}
\int_0^{\lambda^{-Q/2}}\frac{\arctan x}{x}dx.}\label{g}
\end{equation}

In what follows, we always consider (the most interesting from the point 
of view of bandwidth) case $\lambda=1$. It is not, however, difficult
to generalize the argument below to the case of any $\lambda$.
For $\lambda=1$, the expression (\ref{g}) reads:
$\gamma(\varepsilon_{e_i})=4G/\pi Q$. 
In these terms, the Thouless conjecture
says that the total bandwidth behaves as 
$8\gamma(\varepsilon_{e_i})$ asymptotically for large $Q$.
The following more general fact appears to hold and is supported by numerics. 
 
Let $W(x)$ be the total length of the image of $[0,x]$ under the inverse 
of the transform $\sigma=\sigma(\varepsilon)$ (recall that the whole 
spectrum is the image of $[-4,4]$, i.e., the total bandwidth
is $W=W(-4)+W(4)$).
Then for any coprime $P$ and $Q$ (including $P$ changing 
with $Q$ such that $P/Q$ is not small) as $Q\to\infty$:
\begin{equation}
W(\sigma)\sim 4\gamma(\sigma)=\frac{2}{\pi Q}\int_0^{|\sigma|} K(x/4)dx,
\qquad \sigma\in[-4,4].\label{wg}
\end{equation}
The last equation is just another form of (\ref{Le}) for $\lambda=1$;
$K(k)=\int_0^{\pi/2}(1-k^2\sin^2 x)^{-1/2}dx$ is the 
complete elliptic integral of the first kind;
the ``$\sim$'' relation is our conjecture.  

It is easy to derive (\ref{wg}) in the case $P=1$, where we can use 
the known from semiclassics \cite{Watson} asymptotic expression 
for $\sigma(\varepsilon)$.
In this case, the main asymptotic contribution to $W(\sigma)$ comes from 
the bands in a small neighbourhood of $\varepsilon=0$. These bands have width 
of order $1/Q\ln Q$, while the others are exponentially narrow.
For $|\varepsilon|\ll 1$, $P=1$, and $Q\to\infty$,
\begin{equation}
\eqalign{
\sigma(\varepsilon)\sim
4\cosh(\varepsilon Q/4)\cos\left(
\frac{\varepsilon Q}{2\pi}\ln{4Q\over\pi}-\right.\\
\left.2\arg\Gamma\left({1\over 2}+i
{\varepsilon Q\over 4\pi}\right)-{\pi Q\over 2}\right).}\label{d1}
\end{equation}
Hence, the asymptotic contribution of an individual band to $W(\sigma)$ is
\begin{equation}
{2\pi\over Q\ln Q}\arcsin{|\sigma|/4\over\cosh(\varepsilon Q/4)},
\end{equation}
and the number of such bands in 
an interval $dt=d(\varepsilon Q/4)$ is ${2\over\pi}\ln Q{dt\over\pi}$.
Therefore, the sum of all contributions
\begin{equation}
\eqalign{
W(\sigma)\sim{8\over\pi Q}\int_0^\infty\arcsin{|\sigma|/4\over\cosh t}dt=\\
\frac{2}{\pi Q}\int_0^{|\sigma|} K(x/4)dx=4\gamma(\sigma),}
\end{equation}
which completes the derivation of (\ref{wg}) for $P=1$.

If we could show that for a general $P$,  $W(\sigma)\sim c\gamma(\sigma)$,
where $c$ is independent of $\sigma$, then the value of $c$ could be
found using a result of Last and Wilkinson \cite{LW} 
that for any coprime $P$ and $Q$,
$\sum_{i=1}^Q|\sigma'(\varepsilon_i)|^{-1}=Q^{-1}$, where 
$\sigma(\varepsilon_i)=0$, $i=1,\dots,Q$.
This can be written as $|W'_\sigma(0)|=1/Q$ (right or left derivative) 
because, obviously,
$|W'_\sigma(\sigma)|=\sum_{i=1}^Q|\sigma'(\varepsilon_i)|^{-1}$,
where $\sigma(\varepsilon_i)=\sigma$, $i=1,\dots,Q$.
On the other hand, since $K(0)=\pi/2$, we have $|\gamma'_\sigma(0)|=1/4Q$,
which implies that $c=4$.

Putting our results and conjectures together, we can roughly say that
the logarithm of Onsager's partition function, 
the mean of the Lyapunov exponents, and
the asymptotic bandwidth are basically the same object which 
is universal in that for a fixed value of $\sigma$, it does not 
depend on $P$.

I am grateful to P.~Wiegmann for encouraging my interest in the
problem and to R.~Seiler whose valuable suggestions helped me to improve 
this text.
I also thank  L.~Faddeev, J.~Kellendonk, O.~Lipan, A.~Protogenov,
D.~Thouless, and A.~Zvyagin for discussions of this paper and 
valuable comments.

\end{document}